\documentclass[runningheads]{llncs}

\usepackage[T1]{fontenc}
\usepackage[utf8]{inputenc}
\usepackage{graphicx}
\graphicspath{ {./fig/} }

\usepackage[graphicx]{realboxes}
\usepackage{array}
\usepackage{multirow}
\usepackage{longtable}
\usepackage[hidelinks]{hyperref}
\usepackage{booktabs}
\usepackage{svg}
\usepackage{listings}
\usepackage{xcolor}
\usepackage{booktabs}
\usepackage{bbding}

\definecolor{codegreen}{rgb}{0,0.6,0}
\definecolor{codegray}{rgb}{0.5,0.5,0.5}
\definecolor{codepurple}{rgb}{0.58,0,0.82}
\definecolor{backcolour}{rgb}{0.95,0.96,0.96}
\lstdefinestyle{lststyle}{
    backgroundcolor=\color{backcolour},   
    commentstyle=\color{codegreen},
    keywordstyle=\color{magenta},
    numberstyle=\tiny\color{codegray},
    stringstyle=\color{codepurple},
    basicstyle=\ttfamily\footnotesize,
    breakatwhitespace=false,         
    breaklines=true,                 
    captionpos=b,                    
    keepspaces=true,                 
    numbers=left,                    
    numbersep=5pt,                  
    showspaces=false,                
    showstringspaces=false,
    showtabs=false,                  
    tabsize=2
}
\begin{document}

\title{A Cross-Chain Query Language for Application-Level Interoperability Between Open and Permissionless Blockchains}

\titlerunning{Cross-Chain Query Language for Open and Permissionless Blockchains}

\author{Felix Härer\orcidID{0000-0002-2768-2342}} 

\authorrunning{F. Härer}

\institute{Digitalization and Information Systems Group, University of Fribourg, Switzerland
\email{felix.haerer@unifr.ch}\\
\url{https://www.unifr.ch/inf/digits/}}

\maketitle

\begin{abstract}
Open and permissionless blockchains are distributed systems with thousands to tens of thousands of nodes, establishing novel platforms for decentralized applications. When realizing such an application, data might be stored and retrieved from one or more blockchains by distributed network nodes without relying on centralized coordination and trusted third parties. Data access could be provided through a query language such as SQL at the application level, establishing a unified view on application-level data that is verifiably stored. However, when accessing multiple blockchains through their node software and APIs, interoperability cannot be assumed today, resulting in challenges of inhomogeneous data access. In addition, different feature sets and trade-offs exist, e.g., regarding smart contract functionality, availability, distribution, scalability, and security. For increasing interoperability, the paper at hand suggests pursuing the development of a cross-chain query language at the application level. The language abstracts from implementation by providing a standardized syntax, an integrated data model, and a processing architecture for data queries. This research is an extended and updated paper of a prior publication demonstrating the language syntax, data model, and architecture with an evaluation of compatibility against the largest open and permissionless blockchains today. 

\keywords{Distributed Systems \and Blockchain \and Smart Contract \and Decentralized Applications \and Query Language \and Interoperability \and Cross-Chain Swap \and Bridge}

\end{abstract}

\section{Introduction}
\label{sec:introduction}

As of June 2023, a variety of openly accessible blockchains exist with a significant number of active participants. When considering blockchains with at least 1000 daily active addresses, an estimation counts 18 blockchains operating as open platforms for smart contracts or cryptocurrency\footnote{\url{https://www.tradingview.com/markets/cryptocurrencies/prices-most-addresses-active/}, 2023-06-30}. In principle, these platforms can be used for data storage by any business or personal application without centralized coordination and trusted third parties~\cite{belottiVademecumBlockchainTechnologies2019,antonopoulosMasteringEthereumBuilding2019}. Permissionless and verifiable storage are based on algorithmic consensus in contrast to databases and related technologies. In particular, the systems consist of distributed network nodes joining and operating the network at will while any node is able to verify transactions in the blockchain data structure. Decentralized applications are enabled in this way, primarily for programmable money and contracts.

These systems with the components blockchain data, network, and consensus protocol can be considered open and permissionless blockchains (OPB), enabling novel decentralized applications such as programmable money or contracts. Contrary to the distributed systems prevalent in previous decades, well-known OPB now involve the coordinated efforts of thousands to tens of thousands of nodes, forming open and permissionless infrastructures. 

Based on the connectivity of active participants, it is estimated that approximately 16,600 nodes are operating Bitcoin\footnote{\url{https://bitnodes.io/}, 2023-06-30}, 7,600 are operating Ethereum\footnote{\url{https://ethernodes.org/}, 2023-06-30}, and 3,000 are operating Cardano\footnote{\url{https://adastat.net/pools/}, 2023-06-30}. These estimations might not take into account potentially uncounted nodes hidden due to specific configurations, e.g. located behind routers and firewalls. As adoption grows, alongside the increasing number of open and permissionless blockchains, as well as the vast quantities of readily available data, this paper posits the future significance of these platforms for verifiable data storage and execution. Applications interfacing with these platforms encompass various uses, including payments and currency, e-commerce, timestamping, and the attestation of data and web links~\cite{narayananBitcoinAcademicPedigree2017,weberProgrammableMoneyNextgeneration2021,harerDecentralizedAttestationDistribution2022}. 

This research offers an extended and updated study of existing work~\cite{10035048} with the following research problem, objective, and contribution.

\emph{Research Problem.} Software accessing data across open and permissionless blockchains (OPB) today face challenges due to interoperability:

\begin{enumerate}
    \item Inhomogeneous access to data due to various OPB implementations.
    \item Different OPB data models and features exist.
    \item Different OPB trade-offs exist, notably regarding scalability, security, and decentralization.
\end{enumerate}

\emph{Research Objective and Contribution.} The objective of this research is to study the three challenges hindering enhanced interoperability among OPB. The paper contributes a cross-chain query language, established by defining an integrated data model, a grammar and concrete syntax, and a processing architecture. In response to query statements submitted by software applications, data from various blockchain nodes is gathered, integrated into the data model, and processed in accordance with the statements. Considering previously suggested conceptual models and query languages, e.g.~\cite{oliveConceptualSchemaEthereum2020} and~\cite{bragagnoloEthereumQueryLanguage2018}, the language design abstracts from implementation of today's largest OPB. The proof-of-concept implementation demonstrates feasibility and compatibility, but also indicates potential for software to incorporate OPB as integral components of their architecture.

\emph{Application Example.}
Consider a scenario where numerous e-commerce websites participate in shared loyalty programs, issuing reward points for customer purchases. This model is not uncommon among collaboratively operating airlines\footnote{See e.g. \url{https://www.miles-and-more.com/ch/en.html}, 2023-06-30}, among other industries. Given a cross-chain query language, business-level applications across different airlines could access data in a standardized way, re-use queries in their software components, view data on multiple blockchains, integrate and migrate among blockchains, or exchange the underlying blockchains. This is especially advantageous for decentralized scenarios where centralized coordination is limited, e.g. in business networks of different companies relying on separate infrastructure and technology stacks, or generally in decentralized applications.

The paper is organized as follows. Section~\ref{rel} lays out background and related studies. Section \ref{ql} discusses OPB, focusing on their properties essential for the derivation of an integrated data model. The data model, a grammar with a derived concrete language syntax, and a processing architecture follow. A demonstration of feasibility for the language is provided in Section \ref{ev} with a prototype implementation utilizing multiple OPB. Section \ref{co} draws conclusions and provides an outlook.

\emph{Application Example.}
Consider a scenario where numerous e-commerce websites participate in shared loyalty programs, issuing reward points for customer purchases. This model is not uncommon among collaboratively operating airlines\footnote{See e.g. \url{https://www.miles-and-more.com/ch/en.html}, 2023-06-30}, among other industries. Given a cross-chain query language, business-level applications across different airlines could access data in a standardized way, re-use queries in their software components, view data on multiple blockchains, integrate and migrate among blockchains, or exchange the underlying blockchains. This is especially advantageous for decentralized scenarios where centralized coordination is limited, e.g. in business networks of different companies relying on separate infrastructure and technology stacks, or generally in decentralized applications.

The paper is organized as follows. Section~\ref{rel} lays out background and related studies. Section \ref{ql} discusses OPB, focusing on their properties essential for the derivation of an integrated data model. The data model, a grammar with a derived concrete language syntax, and a processing architecture follow. A demonstration of feasibility is provided in \ref{ev} with a prototype implementation utilizing multiple OPB. The final section, Section \ref{co}, draws conclusions and provides an outlook.

\section{Background and Related Work}
\label{rel}

The section at hand introduces blockchain fundamentals, discusses open and permissionless blockchains, and existing interoperability approaches.

\subsection{From Bitcoin to Blockchains}

Following the posting of the Bitcoin whitepaper and corresponding software in 2008 and 2009, respectively~\cite{nakamotoBitcoinPeertoPeerElectronic2008,nakamotoBitcoinSoftwarebasedOnline2009}, the term 'blockchain' emerged as a general term encapsulating its technical architecture. The primary components: (1) a data structure of blocks, arranged in a backward-linked list or any graph, (2) a peer-to-peer network for data distribution, and (3) a consensus protocol, give rise to innovative properties. These notably include the ability to coordinate and validate all operations without trusted third parties or centralized control, open access to all data and operations, and permissionless access, whereby data and operations are not restricted to specific participants~\cite{garayBitcoinBackboneProtocol2015,antonopoulosMasteringEthereumBuilding2019}. Ethereum and subsequent blockchains have enhanced these capabilities by incorporating smart contracts, acting as quasi Turing-complete programs~\cite{woodEthereumSecureDecentralised2022,buterinEthereumNextGenerationSmart2014}. Beyond payments and currency, smart contracts facilitate e-commerce, sales contracts, timestamping, and attestations, among other applications~\cite{ladleifUnifyingModelLegal2019a,narayananBitcoinAcademicPedigree2017,harerDecentralizedAttestationDistribution2022}.

\subsection{Open and Permissionless Blockchains}
\label{opb}

The progressive development and adoption springing from Bitcoin and Ethereum have yielded OPB with diverse characteristics. Table \ref{tab:opb} catalogs five renowned OPB in the order of their public node count, highlighting the properties of their data structures, networks, and consensus protocols, as well as features pertinent to smart contracts.

\begin{table*}[]
\caption{Properties of the largest open and permissionless blockchains by number of nodes.}
\scalebox{0.785}{
\begin{minipage}{\textwidth}
\centering
\begin{tabular}{lllll}
\toprule
Blockchain &
  Data Structure &
  \begin{tabular}[c]{@{}l@{}}Network\end{tabular} &
  Consensus Protocol &
  \begin{tabular}[c]{@{}l@{}}Smart Contract\\Support\end{tabular} \\ \midrule
{[}1{]} Bitcoin\footnote{[18], \url{https://bitnodes.io/}, 2023-06-30} &
  \begin{tabular}[c]{@{}l@{}}Blocks, UTXO \\ data model\end{tabular} &
  \begin{tabular}[c]{@{}l@{}}Bitcoin, approx. 
  \\ 16600 nodes\end{tabular} &
  \begin{tabular}[c]{@{}l@{}}Nakamoto Consensus,\\ 
  Proof-of-Work\end{tabular} &
  \begin{tabular}[c]{@{}l@{}}Stack-based script\\ execution, monetary\\ transactions\end{tabular} \\
{[}2{]} Ethereum\footnote{[29], \url{https://ethereum.org/en/developers/docs/}, \url{https://ethernodes.org/},\\ 2023-06-30} &
  \begin{tabular}[c]{@{}l@{}}Blocks, account \\ state storage in tree \\ data structures\end{tabular} &
  \begin{tabular}[c]{@{}l@{}}Ethereum Mainnet, \\ approx.
  \\ 7600 nodes\end{tabular} &
  \begin{tabular}[c]{@{}l@{}}Gasper (Casper-FFG, LMD- \\ 
  GHOST) Proof-of-Stake\end{tabular} &
  \begin{tabular}[c]{@{}l@{}}Ethereum Virtual \\ Machine, general-\\ purpose programs\end{tabular} \\
{[}3{]} Cardano\footnote{[15], \url{https://adastat.net/pools/}, 2023-06-30} &
  \begin{tabular}[c]{@{}l@{}}Blocks, extended \\ UTXO model\end{tabular} &
  \begin{tabular}[c]{@{}l@{}}Cardano, \\ approx.
  \\ 3000 nodes\end{tabular} &
  \begin{tabular}[c]{@{}l@{}}Ouroboros, \\ 
  Proof-of-Stake\end{tabular} &
  \begin{tabular}[c]{@{}l@{}}General-purpose \\ programs, \\ functional\end{tabular} \\
{[}4{]} Solana\footnote{[30], \url{https://docs.solana.com}, \url{https://solanabeach.io/validators/},\\2023-06-30} &
  \begin{tabular}[c]{@{}l@{}}Block and graph \\ data structures over \\ different time spans\end{tabular} &
  \begin{tabular}[c]{@{}l@{}}Solana Mainnet Beta, \\ approx.
  \\ 1800 nodes\end{tabular} &
  \begin{tabular}[c]{@{}l@{}}Graph-based (proof-\\ -of-history), Proof-of-\\ Stake\end{tabular} &
  \begin{tabular}[c]{@{}l@{}}General-purpose\\ programs\end{tabular} \\
{[}5{]} Avalanche\footnote{[23], \url{https://stats.avax.network/dashboard/network-status/}, 2023-06-30\\} &
  \begin{tabular}[c]{@{}l@{}}Block and graph\\ data structures over \\ different networks\end{tabular} &
  \begin{tabular}[c]{@{}l@{}}Platform/Exchange/\\ Contract (P/X/C) \\ chain, approx.
  \\ 1300 nodes\end{tabular} &
  \begin{tabular}[c]{@{}l@{}}Avalanche (P Chain)\\ Snowman (X/C Chain), \\ Proof-of-Stake\end{tabular} &
  \begin{tabular}[c]{@{}l@{}}Ethereum Virtual \\ Machine (C Chain), \\ general-purpose \\ programs\end{tabular}
\end{tabular}
\end{minipage}
}
\label{tab:opb}
\vspace{0mm}
\end{table*}

\emph{Data Structure.}
The original design of backward-linked blocks in Bitcoin is coupled with additional trees or graphs in most other OPB. Beyond transactional data from blocks, supplementary queries must be performed for non-transactional data or older data that has undergone pruning. For instance, separate tree structures are incorporated for state storage in Ethereum, where balances and smart contract variables can be accessed~\cite{oliveConceptualSchemaEthereum2020}.

\emph{Network.}
Well-known OPB networks comprise approximately 1300 to 16000 nodes. With algorithmic operation and validation, an increased node count augments security, such as in mitigating the risks associated with 51\% attacks and selfish mining~\cite{shrivasDisruptiveBlockchainSecurity2020,saadExploringAttackSurface2020}, which are frequently observed in smaller Proof-of-Work systems, e.g. 'Bitcoin Gold'~\cite{saadExploringAttackSurface2020}.

\emph{Consensus Protocol.} Between 2008 and 2022, a shift in the initially created protocols can be observed, veering away from Proof-of-Work towards Proof-of-Stake, which introduces several trade-offs. While established blockchains such as Bitcoin and Ethereum have prioritized security and decentralization over the years, Cardano~\cite{kiayiasOuroborosProvablySecure2016a}, Avalanche~\cite{rocketScalableProbabilisticLeaderless2020}, and Solana~\cite{yakovenkoSolanaNewArchitecture2018} demonstrated enhancements in efficiency and scalability. This trend is mirrored in the development of novel consensus protocols based on by Proof-of-Stake~\cite{giladAlgorandScalingByzantine2017,ethereumProofofstakePoS2022} with higher efficiency, advantages to environmental impact, enhanced security, and potentially higher distribution and scalability. For example, Ethereum realizes Proof-of-Stake through GASPER, a combination of the consensus algorithm Caspar-FFG ("Casper the Friendly Finality Gadget") and LMD-GHOST (Latest Message Driven Greedy Heaviest Observed Sub-Tree)\footnote{\url{https://ethereum.org/en/developers/docs/}, 2023-06-30}. Based on Caspar-FFG, blocks are proposed in slots of 12 seconds, part of 6.4-minute epochs of 32 slots, by the staking network nodes. The node proposing a block is randomly chosen while other nodes are organized in randomly formed subnets to carry out validations that are aggregated in attestations. Typically, a block is finalized within two epochs with improvements toward single-slot finality\footnote{\url{https://ethereum.org/de/roadmap/single-slot-finality/}, 2023-06-30}. In the case of chain splits, this design together with the fork choice rule has proven itself in practice, demonstrating improved efficiency, decentralization, and security\footnote{\href{https://offchain.medium.com/post-mortem-report-ethereum-mainnet-finality-05-11-2023-95e271dfd8b2}{https://offchain.medium.com/post-mortem-report-ethereum-mainnet- finality-05-11-2023-95e271dfd8b2}, 2023-06-30}. Other blockchains focus especially on scalability, e.g. Solana. However, temporary protocol failures can be observed frequently, resulting in non-availability \cite{haywardSolanaBlamesDenial2021}.

\emph{Smart Contract Support.} Smart contract features are essential for data queries and software applications. Bitcoin offers a limited scripting language employed for programmable monetary transactions and the scalable lightning overlay network. The advent of general-purpose programming in Ethereum and similar platforms introduces a broader range of capabilities and complexity. Currently, most implementations are written and compiled for the Ethereum Virtual Machine, which is present in Ethereum and Avalanche. On the contrary, Cardano and Solana embrace markedly different paradigms. For instance, Cardano supports functional programming, preventing side effects and implementation errors, thus, possibly enhancing security and safety properties\footnote{\href{https://docs.cardano.org/plutus/learn-about-plutus/}{https://docs.cardano.org/plutus/learn-about-plutus/}, 2023-06-30}.

\subsection{Interoperability Between Blockchains}
\label{interop}

Interoperability is widely acknowledged for transactions spanning multiple blockchains, established in cross-chain swaps and similar concepts practically implemented in so-called 'bridges'. Furthermore, efforts towards standardizing inhomogeneous data have commenced not only for query languages.

\emph{Cross-Chain Swaps.} Swaps are typically initiated via a protocol on an originating blockchain, where tokens or arbitrary data are locked to prevent further transfer at the onset. A reciprocal transaction is then issued on a secondary blockchain to the initiator of the cross-chain swap, meaning another party often compensates for the tokens with a different asset on the second chain. This transaction includes a cryptographic proof with a secret that releases the tokens on the initial chain. Finally, the counterparty retrieves tokens from the originating chain. A wide array of protocols and variants have been developed on this foundational principle~\cite{pillaiCrosschainInteroperabilityBlockchainbased2020,shadabCrosschainTransactions2020}. For atomic cross-chain swaps \cite{herlihyAtomicCrossChainSwaps2018a,zakharyAtomicCommitmentBlockchains2020}, atomicity is assured for all transfers involved in a cross-chain swap. Practical implementations in bridges, however, may exhibit different properties and assurances, not necessarily providing atomicity or other guarantees for the completion of the exchange. Bridges are primarily utilized for cryptocurrency exchanges; for example, Multichain\footnote{\url{https://multichain.org/}, 2023-06-30}, Portal\footnote{\url{https://www.portalbridge.com/}, 2023-06-30}, and others\footnote{\url{https://l2beat.com/bridges/tvl\#active}, 2023-06-30} facilitate cross-chain swaps between Ethereum, Avalanche, among others. However, cross-chain swaps and bridges lack standardization and do not provide uniform access or queries.

\emph{Inhomogeneous Data.} Standardization efforts are underway to tackle the issue of inhomogeneous data, with sparse prior work addressing non-uniform access. For Ethereum, one study~\cite{oliveConceptualSchemaEthereum2020} explores a conceptual schema derived from the primary data structures of the blockchain. In \cite{bragagnoloEthereumQueryLanguage2018}, a query language is proposed for the content of blocks and transactions. This language design leans on SQL syntax and supports concepts such as projection and selection within Ethereum. For data analysis, a framework and its implementation based on Scala have been suggested~\cite{bartolettiGeneralFrameworkBlockchain2017}, employing SQL or NoSQL alongside aggregation functions and similar analysis methods.

Another approach~\cite{camozziMultidimensionalAnalysisBlockchain2022} details a data warehouse and ETL process for analyzing Ethereum data using standard SQL with a multi-dimensional data model for attribute dimension queries and data aggregation support. Although this and similar studies might connect to multiple blockchains, they fail to provide homogeneous data access, queries, or simultaneous access to data across multiple blockchains.

Additional work based on SQL includes \cite{liivExplorationStructuredQuery2021}, a study that uses multiple blockchains to populate a standard MySQL database with the third-party service Google BigQuery. However, the reliance on third-party services as data sources presents another commonly observed issue in previous research, where the validation of blockchain data is either impossible or severely restricted. Other methods comprise public connectors between blockchains, blockchains integrating with others, and hybrid approaches~\cite{belchiorSurveyBlockchainInteroperability2021a}.

\emph{Interoperability Limitations in Prior Research.} Present solutions face limitations in terms of (L1.) data access not being homogeneous, (L2.) incompatibility of node software functions and APIs not providing standardized queries, (L3.) software not being able to view and access data on one or more blockchains in parallel, and (L4.) missing verifiability of the blockchain data. The current emphasis is placed on cross-chain swaps and isolated data analysis as opposed to data integration. The query language proposed herein seeks to mitigate these restrictions by suggesting an integrated data model for uniform access (L1.), a grammar and concrete syntax for standardized access (L2.), a processing architecture supporting multiple blockchains in individual queries (L3.) as well as operating multiple nodes locally for verifying transactions (L4.).

\section{Cross-Chain Query Language}
\label{ql}

The following two subsections will detail (A.) the data model and (B.) the grammar with a concrete syntax and a corresponding processing architecture. Query statements are processed as per the architecture delineated in subsection (B.), yielding instances of data model classes using data sourced from the APIs of local blockchain nodes.

\subsection{Integrated Data Model}

\begin{figure*}[htp]
    \centering
    \includegraphics[width=1.00\linewidth]{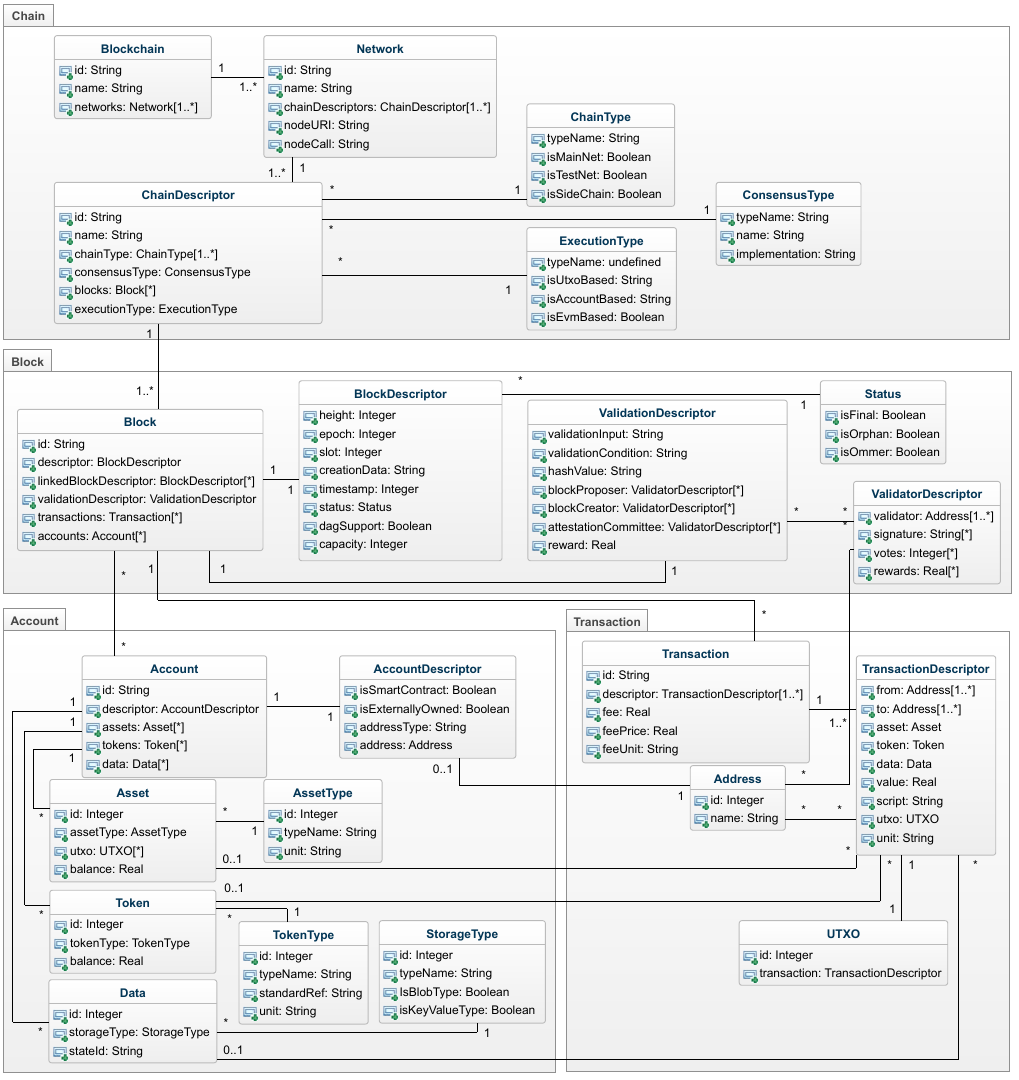}
    \caption{Integrated data model as a UML class diagram differentiating between Chain, Block, Account, and Transaction aspects within the respective packages.}
    \label{fig:data-model}
    \vspace{0mm}
\end{figure*}

The design of the language is predicated on a data model that integrates the principal data structures and attributes of the OPB discussed in Section~\ref{opb}. Building on prior work and existing tools addressed in Section~\ref{opb}, classes and attributes of the five OPB have been identified, generalized, and incorporated into a unified data model. Figure~\ref{fig:data-model} presents the comprehensive data model as a UML class diagram. Table\ref{tab:opb-support} enumerates the main model classes, categorized into four packages to represent the chain, block, account, and transaction concepts of the OPB. The concrete syntax for formulating queries is introduced in subsection~\ref{lang-syntax-proc-architecture}. Statements are articulated in terms of the classes and attributes, specifying the source data using class and attribute names of the data model.

\begin{table*}[ht]
\centering
\caption{Open and permissionless blockchains with related packages of the data model.}
\scalebox{0.79}{
\begin{tabular}{@{}llll@{}}
\toprule
Chain package & Block package & Account package  & Transaction package \\ \midrule

Bitcoin, main chain & Blocks of transactions & - & Values (Bitcoin), UTXO  \\

Ethereum, main chain & Epochs, solots with & Data storage, balances & Values (Ether), \\
                   &  blocks of transactions &   & data incl. tokens  \\

Cardano, main chain & Epochs, slots with     & Addresses, data & Values, assets, \\
                     & blocks for transactions &                 & data, UTXO\\

Solana, main chain & Epochs, slots for transactions & Data storage & Data incl. tokens\\

Avalanche, P/X/C chains & P/X: transaction DAG & P/X: - & X: values, assets, UTXO\\
  
            & C: blocks of transactions & C: Data storage, balances & C: Data incl. tokens
  \\ \bottomrule \\
\end{tabular}
}
\label{tab:opb-support}
\vspace{0mm}
\end{table*}

The concepts of the OPB are shown in the table and data model, encapsulated by the classes of the following packages and classes. Classes of the chain package embody one main network and blockchain for Bitcoin, Ethereum, Cardano, and Solana, as represented by the classes Chain, Network, and ChainDescriptor of the data model. Additional test networks with their distinct blockchains, such as Ropsten and Görli in Ethereum, are represented by the Network and ChainDescriptor classes. In Avalanche, the Network class encompasses one primary network, the first of potentially numerous 'subnets', with separate ChainDescriptor instances for the three P/X/C blockchains.

The Block and BlockDescriptor classes represent blocks, with discrete classes Status for the block's status, ValidationDescriptor for validation via the consensus protocol, and ValidatorDescriptor for the involved validators. Conceptually, blocks across all blockchains are identified by a hash value, supplemented with metadata like timestamps and a height value denoting the block number, assuming no changes to non-final blocks. For instance, in Bitcoin and other blockchains following the original "chain of blocks" concept by Nakamoto~\cite{nakamotoBitcoinPeertoPeerElectronic2008}, a block is linked to its predecessor by a hash value, which is used for validation. This is represented by a Block object with (a) a reference to a BlockDescriptor object, e.g. containing metadata such as the timestamp, (b) a reference to the previous BlockDescriptor object in the linkedBlockDescriptor attribute, and (c) a reference to a ValidationDescriptor object containing the hash value. Regarding non-final blocks in Bitcoin, multiple blocks might be discovered as successors to a given block; however, only one block gets included in the chain, while others are dismissed with an 'orphan' status. In contrast, Ethereum handles similar cases by retaining one block in the main chain while preserving other blocks at the same level with an 'ommer' status. Blocks in Proof-of-Work chains are not explicitly finalized, permitting the assignment of 'orphan' or 'ommer' status to blocks found in parallel to preceding blocks of the chain. Nonetheless, the likelihood of existing blocks being superseded in this manner diminishes over time, as multiple consecutive parallel blocks with greater cumulative work are required. Explicit block finalization, forestalling the emergence of multiple successors, can be observed in more recent Proof-of-Stake blockchains such as Solana.

Concerning data structure, blocks are connected to one or more existing blocks via the linkedBlockDescriptor attribute of the Block class. This connection can establish either a series of backward-linked blocks as mentioned for Bitcoin or a graph structure, such as a Directed Acyclic Graph (DAG) in the Avalanche C chain. The linkedBlockDescriptor relates to the preceding block or, for DAG structures, to any number of previous vertices linked through directed edges. DAG blockchains are indicated by the dagSupport attribute in the BlockDescriptor class, which is set 'true' accordingly.

The representation of blocks further depends on the consensus type. In order to unify the representation for Proof-of-Work and Proof-of-Stake, the ValidationDescriptor class contains generic attributes for storing a hash value, the condition for validation such as the target parameter in Bitcoin and the input, e.g. the Nonce in Bitcoin, as well as attributes for the Proof-of-Stake validation. Here, blocks are proposed, created and verified by attestation involving one or more validators. E.g., a Block in Ethereum is proposed and created by a validator represented as ValidationDescriptor object, and is subsequently verified by attestations. Attestations follow from multiple committees of validators that are represented through the attestationCommittee attribute in ValidationDescriptor, e.g. with multiple multiple addresses and votes. Regarding the creation of blocks, they either contain transactions directly or are grouped into time-based slots and epochs for Proof-of-Stake validation purposes. Upon appending a block, each block or slot undergoes validation, necessitating validators' involvement. As per the ValidationDescriptor class, the creator of a Bitcoin or Ethereum block validates a linked block using the hashValue attribute. Conversely, for other Proof-of-Stake blockchains, block proposers are recorded in the corresponding attributes with attestations, which refer to the ValidatorDescriptor class. Each instance refers to any number of assigned validators who perform attestations of blocks through the committee mentioned before with votes and signatures. Thereby, for Ethereum and other Proof-of-Stake blockchains, the concepts for multiple groups of validators are represented. 

Accounts, a concept prevalent in Ethereum, Solana, and Avalanche, are embedded in blocks to store assets, tokens, or data that are used for smart contracts. For a generic representation of accounts, the data model represents each Account object with an AccountDescriptor object containing the address and an indication whether the account represents a smart contract or an externally owned account of an individual. Concerning account-related data such as assets or tokens, it is important to note that data might represent assets or tokens natively, as seen in Cardano or Solana, or indirectly through data stored within an account. Each account is defined by an ID, with the concept of an address being common to all blockchains. Account storage of assets or tokens can refer to any custom asset or token represented by data in general. For tokens, token standards such as Ethereum's ERC-20 or ERC-1155 are represented by the Token class's attributes. Data storage utilizes binary large objects or key-value stores, which are employed in hash-based mapping data structures.

The concepts of transactions in Bitcoin and Cardano are distinctive due to these blockchains' lack of account structures. Consequently, transactions hold references to unspent transaction outputs (UTXOs) from previous transactions. In this model, a UTXO is included alongside the transferred value and a script that outlines locking conditions or holds data. While data inclusion is implied in Bitcoin, Cardano explicitly accommodates data in transactions and its storage associated with an address for smart contract functionality.

On the other hand, in the case of Ethereum, Solana, and the Avalanche C chain, transactions are stored for the transfer of values, data, assets, or tokens between accounts. In the Avalanche X chain, the transfer of native assets is facilitated through the UTXO model. In the data model, the attributes of Transaction and TransactionDescriptor accommodate transfers between addresses by employing the attributes corresponding to the aforementioned concepts.

\subsection{Grammar and Query Processing Architecture}
\label{lang-syntax-proc-architecture}

The language syntax is rooted in well-established concepts of data query languages, specifically the Structured Query Language (SQL). SQL and similar languages on the one hand permit formalized representation of queries through relational algebra, and on the other hand, allow queries and their execution to be comprehensible to domain experts without deep knowledge of the underlying concepts.

The syntax of SQL is structured around the 'SELECT-FROM-WHERE' block (SFW block). Based on English-language commands, the 'SELECT' clause conducts a projection in the underlying relational model, semantically equivalent to columns. This is followed by the source of the relations in the 'FROM' clause and the selection of tuples utilizing conditions in the 'WHERE' clause. In the relational model, set operations, and notably the Cartesian product, form the foundation for all queries. In particular, this strong theoretic foundation motives the application of the relational model. The relational algebra allows for processing multiple blockchain sources in a SOURCE clause such that (a) data can be combined, e.g., with associative operations, (b) attributes can be selected in a QUERY clause according to highly efficient projection operations, and (c) arbitrary combinations can be produced and filtered in a FILTER clause with very high efficiency. For the cross-chain data language, these concepts are applied in the following manner.

\medskip
\emph{Requirements for Queries.} Query statements consist of query (Q), source (S), and filter (F) clauses as follows:
\medskip

\begin{itemize}
    \item[Q] Query attributes can be any attributes of the data model classes. Each attribute needs to be specified alongside its class, which establishes one column of the query result for each source. This practice prevents ambiguity for conflicting attribute names and allows users to select data based on the required attributes.
    \medskip
    
    \item[S] Sources specify where data is extracted from in terms of blockchain and network classes. This can be paired with additional parameters including specific blocks, transactions, and accounts along with associated assets, tokens, and data.

    To specify each source, attribute values of the identifying attributes from the Chain, Network, and ChainDescriptor classes must be given. This forms the base of the data source from where extraction will begin. Further specificity can be achieved by providing additional classes, attributes, and attribute values of identifying attributes from other classes such as Block, Transaction, Account, Asset, Token, or Data. This level of granularity allows for data queries targeted at one or more blockchains.
    \medskip
    
    \item[F] Filters optionally refine the results of a query based on conditions. By using filters, specific subsets of data can be removed from the query result based on their attributes and attribute values. A filter is specified by a filter function which should contain a comparison operation taking two inputs in the form of query attributes into account. At run-time, filter functions compare the related attributes and their values. Filter functions are applied sequentially to the results obtained before. Due to sequential filtering, the query result only contains data meeting all specified filter conditions. 
    \medskip
\end{itemize}

\begin{figure*}[ht]
    \vspace{4mm}
    \centering
    \includegraphics[width=1.0\linewidth]{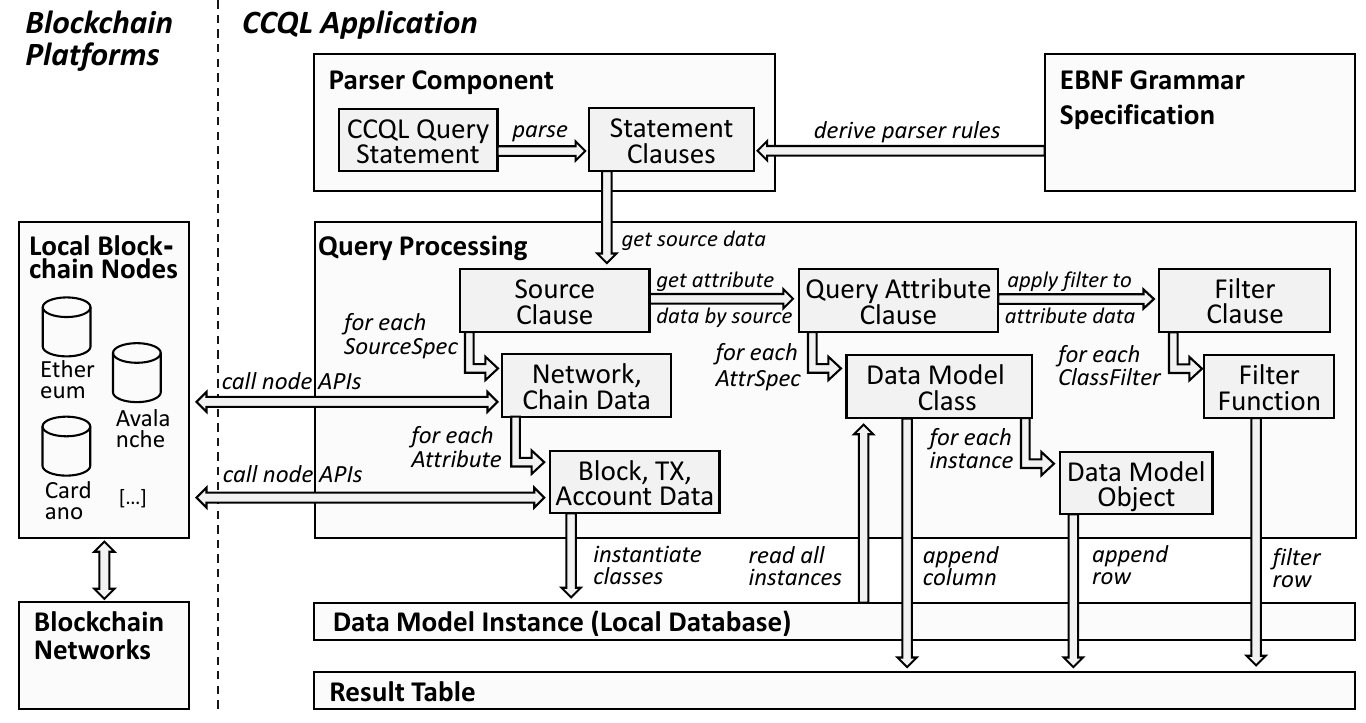}
    \caption{Architecture for application-level query processing.}
    \label{fig:arch-query-proc}
    \vspace{0mm}
\end{figure*}

\medskip
\emph{Grammar and Syntax.} In the provided EBNF (Extended Backus-Naur Form) syntax in Listing~\ref{lst:grammar1}, the structure of a query is divided into a series of clauses. These clauses are used to define the aforementioned aspects of each query and are further detailed and fully specified in the complete grammar\footnote{Available at \url{https://github.com/fhaer/CCQL/tree/main/grammar}}. Query clauses specify projections on the data returned from source clauses, where each source clause relates to the extraction of data as described in the requirements. Finally, filter clauses enable selection by attributes and attribute values through comparison functions. When specifying multiple values within any clause, multiple result sets are the result. In the case of SourceSpec, this would trigger the collection of data from multiple, optionally with a block, transaction, or account, as per requirements and data model. Accounts with assets, tokens, or data are given also according to the specification by the data model. The source and filter clauses are further detailed with the full EBNF grammar specification. For an implementation with a domain-specific language, the concrete syntax might be adopted according to its design guidelines with further usability considerations. 

\vspace{4mm}
\begin{lstlisting}[caption={Excerpt of the grammar in EBNF. Attr: Attribute, Spec: Specification, Val: Value, Desc: Descriptor, I: Instance, Net: Network, Tx: Transaction, Acc: Account.},label={lst:grammar1}]
QueryStatement ::= 
  QueryAttrClause 
  SourceClause
  FilterClause? ";"
QueryAttrClause ::= 
  'Q ' AttrSpec ( ', ' AttrSpec )*
SourceClause ::=
  'S ' SourceSpec ( ', ' SourceSpec )*
FilterClause ::=
  'F ' FilterSpec ( ', ' FilterSpec )*
AttrSpec ::=
  CCQLClass '.' AttrName
SourceSpec ::=
  BlockchainI ':' NetI ':' ChainDescI 
  (':' ( BlockI | TxI | AccI ) )?
FilterSpec ::= 
  CCQLClass '.' AttrName ComparisonFunction IValue
\end{lstlisting}

\medskip
\emph{Query Processing Architecture.} 
Figure~\ref{fig:arch-query-proc} shows the steps involved in query processing within as part of an application architecture. An application initiates the process by issuing query statements to the parser component where clauses are constructed for further query processing in conjunction with a number of connected local nodes. In the query processing component, the source clause is processed for each specific source, i.e., each SourceSpec with network and chain data with their respective attributes leads to the collection of data from the connected nodes. The results are stored as instances of the data model classes. In the next stage of the process, the query attribute clause is processed. Each data model class instance is read to establish a newly appended column in the result table of the query. For the final process stage, the filter clause is applied with each of the specified filter functions, filtering the existing result table. 

\section{Evaluation of Implementation Feasibility}
\label{ev}

The aim of this section is to illustrate the feasibility of implementing the proposed query language with a data model and processing architecture. An implementation compatible with OPB previously introduced has been developed for this purpose\footnote{Available at \url{https://github.com/fhaer/CCQL/tree/main}}. It is composed of two main components: (1.) a language grammar that is formalized using the Eclipse Modeling Framework with a concrete syntax specified on the basis of Xtext\footnote{\url{https://www.eclipse.org/Xtext}, 2023-06-30}. In this way, the syntax is used to derive an external Domain-Specific Language (DSL) with corresponding development and editor environments based on Eclipse. Furthermore, the grammar is implementation-independent and can be re-used in future applications. (2.) a prototype command-line application implementing the language and the data model. The application operates according to the proposed architecture and interacts with nodes of the selected OPB to execute queries. It was developed using Python 3.9 and utilizes the web3.py library to access the OPB\footnote{\url{https://web3py.readthedocs.io/en/stable/}, 2023-06-30}.

\subsection{Software and Hardware Configuration}

Setting up the application involved the following blockchain nodes with a configuration that fully validates all blocks:

\begin{itemize}
    \item Bitcoin node: Bitcoin Core, version 25.0\footnote{\url{https://github.com/bitcoin/bitcoin/}, 2023-06-30}. The initial data synchronization completed after 1 day including the indexing of all transactions.
    \item Ethereum node: Nethermind execution client, version 1.19.0\footnote{\url{https://downloads.nethermind.io/}, 2023-06-30}, together with Nimbus consensus client, version 23.5.1\footnote{\url{https://github.com/status-im/nimbus-eth2/releases}, 2023-06-30} with full validation and execution of transactions. Initial data synchronization completed after approximately 4 weeks.
    \item Cardano node: Cardano node, version 8.1.1\footnote{\url{https://github.com/input-output-hk/cardano-node}, 2023-06-30}. Initial data synchronization completed after approximately 2 days.
    \item Avalanche node: AvalancheGo, version 1.10.3\footnote{\url{https://github.com/ava-labs/avalanchego/releases}, 2023-06-30}. Initial data synchronization completed after approximately 4 days.
\end{itemize}

\begin{figure*}[ht]
    \centering
    \includegraphics[width=1.0\linewidth]{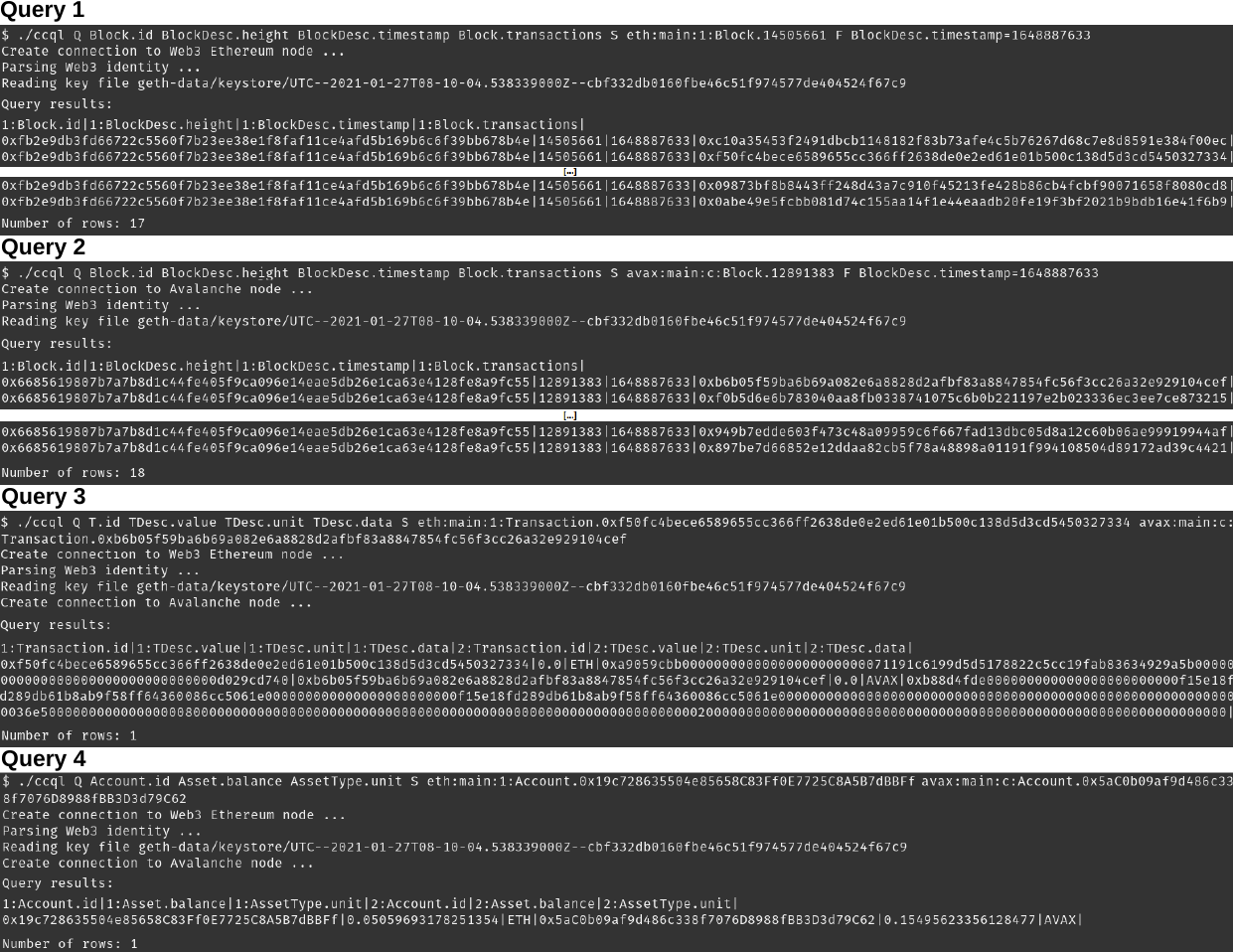}
    \caption{Prototype with query examples for Ethereum and Avalanche showing blocks (Query 1 and 2), transactions (Query 3), and accounts (Query 4).}
    \label{fig:ev1}
    \vspace{0mm}
\end{figure*}

Accounting for typical application scenarios related to businesses or individuals, the data synchronization was carried out on a consumer-grade laptop in all cases. The laptop was equipped with AMD Ryzen 7 5700U CPU, 16 GB RAM, and SK Hynix BC711 NVMe SSD running Ubuntu 22.04. For the synchronization, the laptop was continuously connected to a 1 Gbit/s fiber internet connection. 

To establish feasibility, query statements were evaluated using the prototype, which will be elaborated in the subsequent section. Each statement was executed on the laptop and the locally running blockchain node software without the involvement of further web services or APIs, realizing the processing architecture. As the node software was fully validating and storing data locally, it enabled the generation of query results without network access. It follows that query performance is independent of network latency and solely constrained by the local CPU and IO performance of the device at hand.

\subsection{Query Processing}

The prototype application, realizing the architecture illustrated in Figure~\ref{fig:arch-query-proc} as described in Section~\ref{lang-syntax-proc-architecture}, was used to evaluate typical queries as follows.

The first query example, illustrated in Figure~\ref{fig:ev1}, shows the task of identifying transactions within a block. Here, the query attributes define the Block and BlockDescriptor (BlockDesc) classes along with the properties of the block ID, Height, Timestamp, and transactions. The following source clause specifies Ethereum, the main network, chain 1, and a block number. The query terminates by applying a filter to the timestamp attribute. The output of the query manifests as attributes prefixed by the source number, each displaying instance-level data from the data model with corresponding values. For instance, Block.id and Block.height are denoted as '0xfb2e[...]' and '14505661', respectively. 

Given a specific block, query attributes could be added for continuing the investigation throughout the data model, e.g. identifying accounts in blocks using Block.accounts followed by Asset.balance in order to retrieve their balances. In a cross-chain scenario, a corresponding transaction might be located on another blockchain, e.g. transferring assets or data. By specifying a block and timestamp, transactions occurring with the same or similar timestamp might be queried. In the second query of Figure~\ref{fig:ev1}, this example can be seen with the aforementioned classes and attributes. Investigating this scenario further based on the Ethereum and Avalanche transactions, transactions occurring at the same timestamp in both blockchains were located and queried in the third query displayed in Figure~\ref{fig:ev1}. For obtaining asset transfers and data, the source clause specifies Transaction (T) and TransactionDescriptor (TDesc) classes with attributes for value and data, respectively. The source is addressed by hexadecimal transaction IDs on the two blockchains. From the query results, it can be observed that both transactions are data transactions, transferring assets of value 0.0 and data represented in hexadecimal format. For investigating assets, accounts might be queried in addition, for example as demonstrated in Query 4 of Figure~\ref{fig:ev1}. Given the transactions with matching timestamps together with the involved assets here indicates the exchange of tokens in a cross-chain swap scenario. 

\subsection{Discussion}

The prototype provides uniform data access to OPB, such as the retrieval of asset and data transfers across multiple blockchains. As per the defined grammar, data access is standardized, facilitating statements that involve one or more blockchains. For meaningful utilization of blockchain properties, it is essential to operate blockchain nodes locally, which can involve significant time and cost for initial synchronizations.

The architecture of the data model follows a data integration approach, where data conforming to well-known OPB can be stored by populating pertinent classes. In contrast, merely relying on multiple individual data models would fail to address the issue at hand.

Due to the generic and unifying approach, the prototype in its current form has limited support for advanced concepts specific to individual OPB. For instance, calculating transaction fees involving additional utility tokens falls outside the scope of this model. Functionality-wise, limitations concern the processing of queries with filters. Currently, the prototype is constrained to the sequential application of filters with equality comparisons. In addition, the processing is limited to the classes and attributes present in the data model. In future development, further blockchain-specific processing logic and attributes could be added to the architecture and the data model, respectively. At this point, the prototype demonstrates the general feasibility of a domain-specific language that supports multiple blockchains. Regarding atomicity, it is supported on a technical level for query transactions within the data model, however, the API-based access approach does not allow for atomicity guarantees that depend on the software node implementations of different blockchains. Oftentimes, APIs do not provide specific atomicity or strong consistency guarantees and instead rely on eventual consistency. For this reason, access through APIs limits the possibility of strong guarantees for the prototype.

\medskip
\section{Conclusion and Outlook}
\label{co}

This paper presents a cross-chain query language grammar, data model, and processing architecture aimed at facilitating uniform data access across multiple blockchains. The approach enables homogeneous data access, query standardization, addressing multiple blockchains within individual queries, and local validation of blockchain data. These facets were only partially covered in previous research.

The feasibility of implementing the language with its processing architecture has been positively evaluated using a prototype, despite the functional limitations mentioned in the previous discussion. Using the proposed approach of application-level interoperability, software can leverage multiple blockchains to establish a unified view on data while relying on verifiable transactions that are part of an open and permissionless infrastructure. 

In future research, these concepts can serve as a basis for addressing further integration aspects among blockchains, e.g. in terms of augmenting data storage distributed on multiple blockchains, and provide advanced integration methods towards enabling blockchains as decentralized application platforms.

\section*{Acknowledgment}

This work is supported by the Swiss National Science Foundation project Domain-Specific Conceptual Modeling for Distributed Ledger Technologies [196889].

\bibliographystyle{splncs04}
\bibliography{references}

\end{document}